\documentclass[%
aps,%
prb,%
twocolumn,
superscriptaddress,%
showpacs,%
preprintnumbers,%
byrevtex,%
floatfix%
]{revtex4}

\usepackage{ifthen}
\usepackage[usenames]{color}
\usepackage{amssymb}
\usepackage{amsfonts}
\usepackage{amsmath}
\usepackage[dvips]{graphicx}

\def\ii{\textrm{i}\,\!}

\newcommand{\beann} {\begin{eqnarray*}}
\newcommand{\eeann} {\end{eqnarray*}}
\newcommand{\bea} {\begin{eqnarray}}
\newcommand{\eea} {\end{eqnarray}}

\newcommand{\lsb} {\left[}
\newcommand{\rsb} {\right]}
\newcommand{\lrb} {\left(}
\newcommand{\rrb} {\right)}
\newcommand{\lcb} {\left\{}
\newcommand{\rcb} {\right\}}
\newcommand{\lab} {\left\langle}
\newcommand{\rab} {\right\rangle}

\newcommand{\vect} {\mathbf}

\newcommand{\Ham} {{\mathcal H}}

\newcommand{\Cond} {G}

\newcommand{\g}{\,^\circ}
\newcommand{\ie}{\textit{i.e.~}}

\begin{document}
\date{January 17, 2006}

\title{Scaling of the conductance in gold nanotubes}

\author{Miriam del Valle}
\affiliation{Departamento de F\'{\i}sica Te\'{o}rica de la Materia Condensada, Universidad
Aut\'{o}noma de Madrid, Facultad de Ciencias, C-V, E-28049 Madrid, Spain}
\affiliation{Institute for Theoretical Physics, University of Regensburg, D-93040 Regensburg, Germany}
\author{Carlos Tejedor}
\affiliation{Departamento de F\'{\i}sica Te\'{o}rica de la Materia Condensada, Universidad
Aut\'{o}noma de Madrid, Facultad de Ciencias, C-V, E-28049 Madrid, Spain}
\author{Gianaurelio Cuniberti\thanks{Corresponding author}}
\affiliation{Institute for Theoretical Physics, University of Regensburg, D-93040 Regensburg, Germany}

\begin{abstract}
A new form of gold nanobridges has been recently observed in
ultrahigh-vacuum experiments, where the gold atoms rearrange to
build helical nanotubes, akin in some respects to carbon nanotubes.
The good reproducibility of these wires and their unexpected
stability will allow for conductance measurements and make them
promising candidates for future applications . We present here a
study of the transport properties of these nanotubes in order to
understand the role of chirality and of the different orbitals in
quantum transport observables. The conductance per atomic row shows
a light decreasing trend as the diameter grows, which is also shown
through an analytical formula based on a one-orbital model.
\end{abstract}

\maketitle
Gold is known to be a good conductor for ages but its use in
nanoelectronics delivers unexpected behaviors differing from its
bulk properties. Recent
experiments~\cite{KondoT97,KondoT00,OshimaOT03} indicate that gold
nanowires exhibit fascinating ordered structures that resemble those
of the by now well-known carbon nanotubes (CNTs).~\cite{ReichTM04}

High resolution microscope images show that these wires consist of
coaxial nanotubes of helical atom rows coiled around the wire axis.
These wires present therefore different possible helicities as CNTs
do, depending on the angle in which the atom rows rotate around the
tube axis. Experimental evidence so far seems to support the fact
that small chiral angles are preferred by the fabricated nanotubes.
The multi-shell nanotubes keep the difference in the number of atom
rows between adjacent shells constant, leading to a kind of magic
shell-closing number which is always seven. This fact can be
understood as a new inner shell arises when the radii differ in
about $a = 2.88\ \textrm{\AA}$, the neighboring distance of bulk
gold, which implies a perimeter difference ($2\pi a \sim 7a$)
corresponding to seven atom rows nearly parallel to each other. So
single-walled nanotubes should exist with six or less atom rows, and
actually one single-walled tube with five atom rows has been
observed.~\cite{OshimaOT03} With seven or more atom rows a
double-walled tube should be seen, which becomes a triple-walled
nanotube when more than fourteen rows are present, and so on. The
good reproducibility of these wires, and their
stability,~\cite{GuelserenET98,Bilalbegovic98,daSilvaSF01,TosattiPKCT01}
which is better for those with outer tubes with an odd number of
atom rows, makes them promising candidates for future applications.
Ballistic conductance measurements will lead to a confirmation of
the proposed structural model through its derived electronic
structure. Astonishingly enough several works hint at the fact that
the conductance per atom row might decrease with the number of
circumferential atoms.~\cite{SengerDC04,OnoH05} In this Letter, we
provide a theoretical understanding of this phenomenon by means of
both numerical and analytical calculations.

\begin{figure}[b]
\centerline{\includegraphics[width=.99\linewidth]{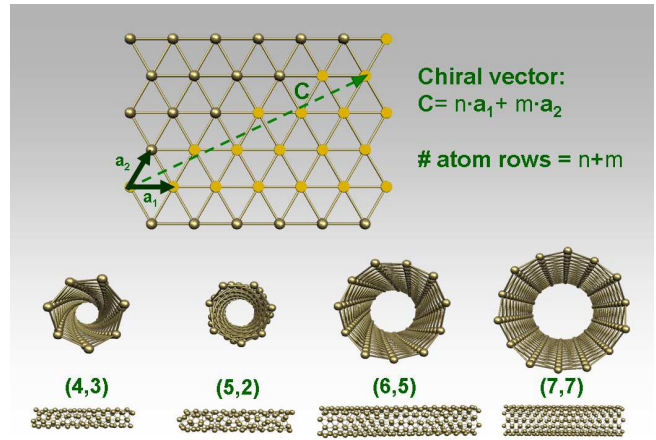}}
\caption{\label{fig_tubes}
Some gold nanotubes are shown as well as the 2D triangular network,
from which we can build these models. The coordinate system used is
sketched on the lattice, being $\vect{a}_1$ and $\vect{a}_2$ the basis vectors,
and $\vect{C}$ the chiral vector specifying the nanotube and corresponding to
its circumference. The highlighted atoms on the lattice are those defining the
region of inequivalent chiral vectors, which spans an angle of $30\g$ as can be easily
understood analyzing the symmetry of the lattice.%
}
\end{figure}
Determining the structure of the nanotubes is fundamental in order
to understand their physical properties, such as conductance
quantization. Experimental evidence points out that a structural
transition from the thicker, internally crystalline wires into these
thinner regular but noncrystalline ones takes place at a critical
radius below $2$ nm. Lattice spacing has been measured to be almost
$2.88\ \textrm{\AA}$, the neighboring distance of gold, for all
these wires with diameters between $0.5$ nm and $1.5$ nm. We
therefore take this distance as the interatomic spacing of the
triangular lattice, as we can think of these cylindrical tubes as
rolled-up lattice planes of fcc Au (111), as depicted in
Fig.~\ref{fig_tubes}. Atoms in successive shells appear to strain a
bit to maintain commensurability between inner and outer shells.
Ab-initio calculations show though that this shear strain should
have just a very small effect on the electronic
structure.\cite{YangD05}

Despite the strong intershell interactions, these structures resemble
greatly those of CNTs, where the honeycomb network of carbon atoms is
replaced by the triangular lattice of gold atoms. These are actually
complementary networks, and the gold nanotube structure can be
obtained from the one of CNTs by putting gold atoms at the center of
honeycomb framework.

\begin{figure}[t]
\centerline{\includegraphics[width=.99\linewidth]{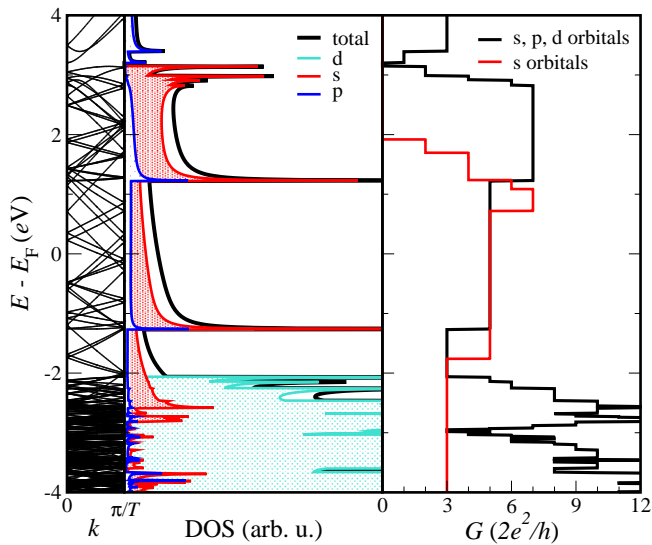}}
\caption{\label{fig:43}
Onedimensional band structure, density of states and conductance for
the $(4,3)$ gold nanotube. For this tube the translational vector is
$T = |\vect{T} | = 3.03$ nm. The energies are shifted to set the
origin at the Fermi energy, as in the next figures.
}
\end{figure}

We use the indices $(n,m)$ to classify the gold nanotubes (AuNTs),
where $n$ and $m$ are integers defining the chiral vector, which is
wrapped to form the tube. The chiral vector is then $(n \vect{a}_1 +
m \vect{a}_2)$, perpendicular to the tube axis, where the vectors
$\vect{a}_1$ and $\vect{a}_2$ span the 2D unit cell. The number of
atom rows coiling around the tube axis is then $n+m$ while the
nanotube translational vector reads $\vect{T} (n,m) = \lrb (2m +
n)/d_R\rrb \vect{a}_1 - \lrb (2n + m)/d_R\rrb \vect{a}_2$, being
$d_R$ the greatest common divisor of $(2m + n)$ and $(2n + m)$, as
$\vect{T}$ should be the smallest lattice vector in its direction.
Due to the symmetry of the triangular lattice the chiral vectors of
all inequivalent tubes are comprised in a $30\g$ angle, that is, in
a one-twelfth irreducible wedge of the Bravais lattice. We can then
compromised and use only indices with $n>0$ and $0<m<n$. The tubes
of the form $(n,0)$ and $(n,n)$ are achiral, presenting thus no
handedness. Only wires with an even number of strands may have the
structure $(n,n)$.~\cite{notation}

For a fixed $n$, the radius is minimized for higher values of $m$
(tension decreases with shrinking radius), so as to fulfill
simultaneously tightest external packing and minimal wire
radius.~\cite{TosattiPKCT01} The largest $m$ for odd number of atom
rows is $n-1$, leading always to a finite chirality and thus a
helical structure in these cases. The outer tubes with an even
number of atom rows seem to take non-helical structures with shorter
periods. The helicity of the tubes is observed experimentally
through a wavy modulation of the STM
images.\cite{KondoT00,OshimaOT03}

\begin{figure}[t]
\centerline{\includegraphics[width=.99\linewidth]{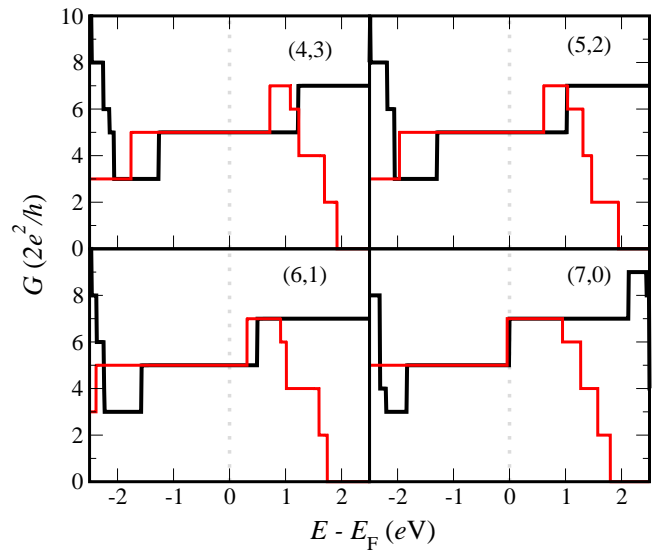}}
\caption{\label{fig:7rows}
Conductance of AuNTs with seven atom rows but different chiralities.
Red lines indicate results including $s$ orbitals only, whereas the
black
ones are showing the calculations with all nine outer orbitals.%
}
\end{figure}

We study the transport properties of several of these AuNTs using a
multiorbital tight-binding Hamiltonian and Green function
techniques. In particular we adopt a
Slater-Koster-type~\cite{SlaterK54} tight-binding  approach, with
the parameters taken from the nonorthogonal parametrization of
Papaconstantopoulos and co-workers.\cite{Papaconstantopoulos86}. Our
approach remains of microscopic nature since the symmetries of the
atomic $s, p,$ and $d$ valence orbitals are taken into account. Ab
initio studies including shell effects could be needed for greater
quantitative accuracy.

As a result we analyze the differences in the transport
calculations that raise from the inclusion of all outer \textit{s},
\textit{p} and \textit{d} orbitals instead of just the simplified
model of \textit{s} orbitals, which is a tempting approximation for
these kind of systems and has been used in other
works.\cite{OnoH05} In order to compare results, we will thus use
both a one-orbital Hamiltonian describing just the $s$ orbitals, and a
nine-orbital Hamiltonian, containing $s, p$ and $d$ orbitals including
all conduction electrons. In doing that we restrict our calculations to
nearest-neighbors interactions.

The Hamiltonian describing our systems can be written as

\bea \Ham = \sum_{\lab i,j \rab,\alpha,\beta } H_{i\alpha,j\beta}
c_{i\alpha}^\dagger c_{j\beta}^{\phantom{\dagger}} \eea

where the indices $i,j$ run over the atom sites and the indices
$\alpha,\beta$ mark the different orbitals. The transfer integrals
$H_{i\alpha,j\beta}$ are the on-site energies or hopping
parameters, dependent on whether $i$ equals $j$ or not.

The conductance is calculated from the transmission values applying
the Landauer formula, using Green function
techniques.~\cite{CunibertiGG02} In particular, we derive the
elastic linear response conductance via the Fisher-Lee formula for
the quantum mechanical transmission~\cite{FisherL81}:
$G=\frac{2e^2}{h} \textrm{Tr}\lcb \vect{\Gamma}_{\rm L} \,
\vect{{\cal G}}  \, \vect{\Gamma}_{\rm R} \, \vect{{\cal G}}^\dagger
\rcb,$ where $\vect{\Gamma_{\rm L/R}} = \ii \lrb \Sigma_{\rm L/R} -
\Sigma_{\rm L/R}^\dagger \rrb$ being $\Sigma_{\rm L/R}$ the self
energy of the left or right lead respectively, and where
$\vect{{\cal G}}$ is the Green function of the central region
dressed by the electrodes.

We consider semi-infinite single-walled nanotubes (SWNTs) as leads.
Considering the leads as bulk gold will yield an increased imaginary
part of the lead self energies, which will slightly decrease the
conductance values. For multi-walled nanotubes (MWNTs) it has been
suggested that the conductance $G$ is mainly determined by the outer
shell which is also responsible for the structural
stability.~\cite{HasegawaYH01} A slight reduction of the conductance
is expected from the finite size effects.


The calculated density of states and the conductance for the gold
nanotube $(4,3)$ are shown in Fig.~\ref{fig:43}. This tube is one of
the simplest possible candidates with seven atom rows around its
circumference (Fig.~\ref{fig_tubes}). The colored areas show the
contribution of the different orbitals to the total density of
states, indicating a dominance of $s$ orbitals around the Fermi
level. Considering only $s$ orbitals around the Fermi energy is thus
a good approximation, but when applying a bias it can be critical to
include all $s, p, d$ outer orbitals, as they play an important role
in the opening of new conduction channels.

\begin{figure}[t]
\centerline{\includegraphics[width=.99\linewidth]{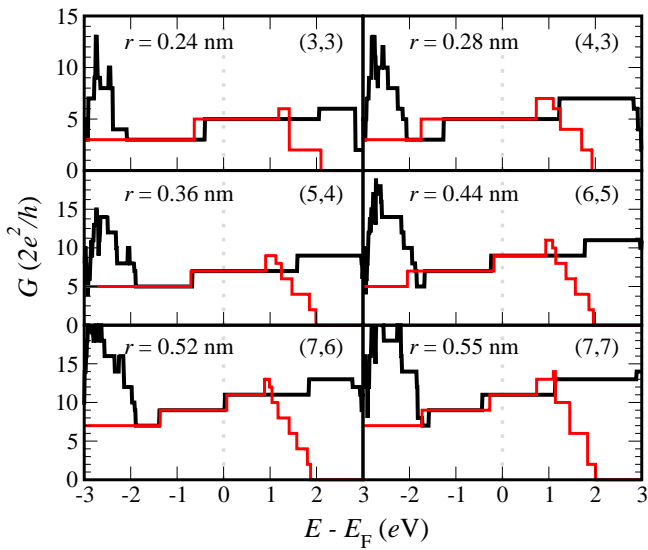}}
\caption{\label{fig:diffRows}
Conductance of AuNTs with different radii $r$. Red lines indicate
results including $s$ orbitals only, whereas the black ones are
showing the calculations with all nine outer orbitals.
}
\end{figure}

A comparison of the conductance of nanotubes of the same diameter
but different chiralities, as shown in Fig.~\ref{fig:7rows},
suggests that no significant difference is expected in their
conductance values at the Fermi energy. For applied bias voltages
the current will show the difference in energies at which new
channels start taking part in transport.

In Fig.~\ref{fig:diffRows} the conductance is plotted for nanotubes
of different radii, chosen as to have the smallest chirality
possible for a given radius. The number of atom rows around the
nanotube axis in each of them is $6, 7, 9, 11, 13$ and $14$
respectively. We see that, though the conductance increases for
thicker nanotubes, this is not the general trend if divided by the
number of atom rows. We can observe again how the expansion of the
orbital basis to include all conduction electrons changes the
energies at which new channels open, being critical to obtain
correct current values.

\begin{figure}[t]
\centerline{\includegraphics[width=.89\linewidth]{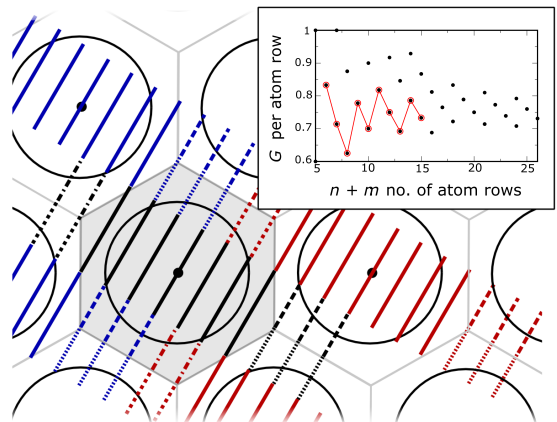}}
\caption{\label{fig:fig5}
Schematic illustration of the reciprocal lattice of the 2D hexagonal
Bravais lattice of Au (111) where the first Brillouin zone is
highlighted through a darker background. The $k$-vectors lying at
the Fermi surface give rise to a nearly perfect circle in this 2D
structure. The black cutting lines represent the Brillouin zone of a
nanotube in an extended zone scheme which allows the use of the
dispersion relation calculated for the 2D gold layer to get the
bands of a folded nanotube. In this example we show the wave vectors
giving the bands in a (6,0) AuNT, a good candidate for this
visualization as it has a small number of bands. These lines are
plotted as solid bands inside the first Brillouin zone, and colored
when folded back into it through the reciprocal lattice vectors of
the gold layer. In the inset the conductance per atom row at the
Fermi energy based on the analytical model accounting only for
$s$-orbitals described in this Letter is shown for all possible
($n$,$m$) NTs, reproducing exactly the values of the numerical
calculations.
}
\end{figure}

By analyzing the Brillouin zone of the two-dimensional gold lattice
taking into account only $s$ orbitals, one can easily demonstrate
after studying its dispersion relation that the Fermi surface is
with great accuracy approximated by a circle with a radius close to
$k_{\rm F}= 2 \sqrt{\pi}/(\sqrt[4]{3}\,a)$, where $a = 2.88\
\textrm{\AA}$. This simple model allows us to have an analytical
approach to the conductance, as the number of energy bands crossing
the Fermi surface, since the Brillouin zone of the gold nanotubes
consist of parallel line segments due to the quantization of the
wave vector along the circumferential direction. For simplicity we
can restrict this counting to the first Brillouin zone by folding
back to this area the segments lying outside it, which will elongate
the lines inside it, as pictured in
Fig.~\ref{fig:fig5}.~\cite{Harrison60} The conductance can then be
written as

\bea \Cond = \frac{2 e^2}{h} \lsb 2\, \textrm{Int}\lrb
\sqrt{\frac{n^2+m^2+nm}{\pi \sqrt{3}}}\rrb + 1 \rsb \eea

and is only dependent of the nanotube indices and not on hopping
parameters or on-site energies. This models yields a perfect
matching with the numerical calculations described before. The
result of this approximation of the conductance at the Fermi level
is plotted in the inset of Fig.~\ref{fig:fig5} for nanotubes of
different radii and chiralities, presenting the conductance per atom
row. The solid lines in this figure put an accent on the results of
AuNTs with the smallest helicity as characterized in
Fig.~\ref{fig:diffRows}. We can observe how the envelope of all
these points slowly decreases with increasing number of coiling atom
rows, or likewise with increasing diameter. The conductance values
will nevertheless be reduced by considering more realistic gold
leads. This trend has been also found in a first-principles
one-orbital calculation.~\cite{OnoH05}

\begin{figure}[t]
\centerline{\includegraphics[width=.89\linewidth]{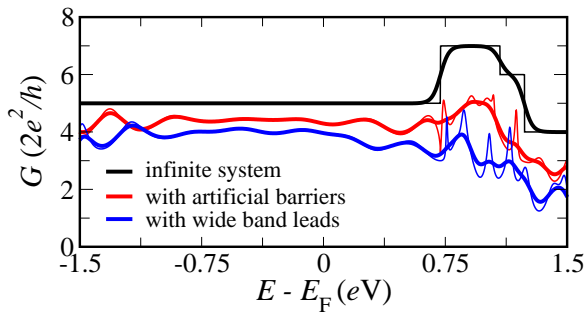}}
\caption{\label{fig:mesos}
Conductance of the ($4$,$3$)-AuNT for a perfect infinite NT (black
line), for a finite tube with barriers in the connecting regions to
the electrodes (where the constant $\alpha$ introduced in the text
equals $0.6$) and for a finite tube like in the previous case but in
the wide band limit approximation taken at the Fermi energy. The
colored lines show oscillations due to scattering at the ends of the
finite size tube. The thicker lines correspond to room temperature
conductance.\cite{BagwellO89}
}
\end{figure}
Gold nanotubes have been observed as short ($L \sim 4$ nm)
conductors between bulk gold electrodes. This calls for an extension
of our theory to account for finite size effects. We did this by
parametrically introducing into the nanotube two tunneling barriers
at a distance $L$. Fig.~\ref{fig:mesos} shows the conductance
between two homologous semiinfinite electrodes which are taken into
account through modified self energies. This modification is
introduced by a multiplicative factor $\alpha$, which simulates the
barriers. The finite size effects can be equally made apparent by
considering wide band leads, \ie $\vect{\Sigma}_{{WB}} (E) \cong
-\ii \,\mathrm{Im}\vect{\Sigma}(E=E_{\textrm F}$). As we can see in
Fig.~\ref{fig:mesos}, we obtain the same kind of oscillating
behavior with both approximations. This shows that the results of
our model provide an upper limit for the conductance.

In conclusion, we investigated the electronic properties of several
gold nanotubes. We built the Hamiltonians of the systems by applying
a tight-binding model for $s$ orbitals as well as for all outer
orbitals. All AuNTs are equally metallic independently of their
chirality or diameter in contrast to the results obtained for their
carbon counterparts,~\cite{SaitoFDD92} but expected from gold
electronic nature. In the case of gold the quantization of electron
waves along the circumference of the tube results in a Brillouin
zone composed of line segments that always cut the Fermi surface of
the Au [111] layer due to the continuity of this surface.

The $s$-orbital calculations demonstrate that a one-orbital approach
to the problem of gold nanotubes is not a bad approximation, but
should be avoided when applying a finite bias voltage to the system.

An analytical formula for the conductance of the $s$-orbital
calculations, matching perfectly the numerical results, shows that
there is a slight decrease of the conductance per atom row as the
radius increases.


\begin{acknowledgments}
This work was partially funded by the Volkswagen Foundation
(Germany) under grant No.~I/78~340, by the MEC (Spain) under
contracts MAT2005-01388, NAN2004-09109-CO4-04, and by the CAM under
contract No. S-0505/ESP-0200. MD acknowledges the support from the
FPI Program of the Comunidad Aut\'{o}noma de Madrid.
\end{acknowledgments}


\begin{thebibliography}{22}
\bibitem[{not()}]{notation}
\bibinfo{note}{We can decode our notation $(n,m)$ into the one used in
  parallel with lattice vectors spanning an angle of $120\g$ like $(n+m,n)$ for an equivalent AuNT for transport.}
\end{thebibliography}

\end{document}